# Tuning plasmon excitations in pure and transition metal-doped arrays of noble metal nanochains

Neha Nayyar, Volodymyr Turkowski, Talat S. Rahman*

Department of Physics, University of Central Florida, Orlando, FL -32816, USA

*Corresponding Author, e-mail address: Talat.Rahman@ucf.edu

KEYWORDS. Nanoplasmonics, Nano-chains, TM-doped chains

ABSTRACT. We study the plasmonic properties of coupled noble-metal nanochains in the case of different number of coupled chains and doping by different transition-metal (TM) atoms within the time-dependent density-functional theory (TDDFT) approach. We find that as the number of chains in the array increases the plasmon peak shifts from the sub-eV towards the visible range. As doping with TM atoms increases, the visible absorption band broadens, owing to formation of additional plasmon peaks. The optical response is very sensitive to the type of doped atoms, their number and position; in particular, the additional peaks are most pronounced in the case of weak doping when they correspond to local plasmon oscillations around the impurity atom. These effects have a potential to be used in various modern technologies, from sensors to solar cells. Most of the studies of nano-plasmon effects have been focused on alkali- and noble-metal systems with extended s-electron states, while it was believed that doping with TM atoms with their more localized charge as a rule leads to an attenuation of the plasmon modes. We demonstrate that TM atoms can play a constructive role in plasmon generation in small chain systems, and that plasmonic modes can emerge even in some pure TM nanochains.

**INTRODUCTION**

Optical properties of nanosystems are highly sensitive to their size, shape and the chemical composition, and can dramatically differ from those of their bulk cousins. There are many properties of these systems, which can be used for practical applications, like linear chains of nanoparticles which have been proposed as possible conduits for electromagnetic energy transport (optical interconnects) (see, e.g., Ref. 1-4). In order to be able to tune the system properties, including the absorption spectrum, it is essential to understand the dependencies of its properties on the shape, size, composition and other parameters. Over the past decade, a wide variety of plasmonic gold and silver structures has been fabricated to manipulate the light absorption at the nanometer scale for novel applications and basic research.[5-6] These applications rely to some extend on the ability to tune the particle plasmon resonances, which has played a crucial role in stimulating the current interest in nanoplasmonics.

One of the first theoretical studies on the collective excitations in a few-atom clusters were carried out by Kummel et al.[7] who showed that collective excitations exist even for very small clusters. More recent work addresses the mechanism of the collective excitations in the alkali-metal clusters.[8-10] Ma et al.[11] studied the sensitivity of plasmon resonance in Au nanoparticles and their dimers as a function of the particle size and the inter-particle distance. Studies of the plasmon excitations in planar Na structures[12-13] reveal the importance of dimensionality in the formation and development of the plasmon peaks. Recently, Scanning Tunneling Microscope (STM) experiments have demonstrated development of 1D band structure in Au chains on NiAl(110) when the number of atoms in the chain exceeds 10.[14] Inspired by this experimental finding, theoretical calculations have also predicted the presence of collective plasmon modes in a few-atom chains of several metallic elements: Na,[15] Ag,[16] and Au[17,18] (for over-review, see, e.g., Ref. 32). Experimental observation of such collective excitations requires the chains to be grown on a substrate that does not quench them rapidly. While there are theoretical indications that NiAl(110) surface does not affect the electronic properties of the Au chains,[19] its metallic nature precludes a short lifetime for any plasmon excitation. On the other hand, it is possible to grow Au chains and wires on semiconductor substrates such as Si (557)[20], Ge(001)[21] and quartz[22] which may be amenable for capturing the plasmon effects (especially with the bandgap much larger than the chain-plasmon and other excitation energies of interest). In principle, one may expend a significant change of the interatomic distances when the chains are put on the substrate. On the other hand, as our calculations demonstrate, the qualitative results remain qualitatively and semi-quantitatively the same when the bondlengths, one of the most important effects of the substrate, change (in a reasonable, ~0.01-0.1nm, interval).

In our previous study,[23] we examined the role of TM doping in the generation of plasmon modes in single Au chains, and found that it leads to several changes in the absorption spectrum, including (most strikingly) the appearance of a new local mode. These new excitations are a result of a complex rearrangement of the ionic potential around the impurity atom felt by the delocalized electrons participating in the plasmon oscillations around the TM atom, as well as a collective effect of the "localized" (TM d-) and de-localized (s) electrons. Indeed, the mutual effects of both electronic subsystems may be nontrivial, in particular leading to a change of the spectral function of the localized electrons (see, e.g., Ref. 24), or to local electronic resonances around the dopant atom, as was shown experimentally in the case of Pd-doped Au chains on NiAl(110).[25] This complexity opens the door to new opportunities for tuning the optical properties of noble metal nanostructures by changing the TM doping. On the other hand, one may also tune the optical spectrum of the system by changing its size, shape and geometry to find the arrangements that may lead to the absorption of visible light. In this study, we consider one type of such a system: an arrangement of small chains of (different types of) atoms in one or several arrays. We analyze the optical properties of these arrays as a function of their elemental composition and number of chains in the array for several noble-metal chains, both in pure form and doped with TM atoms, and also study the absorption spectrum when two pure chains of noble and TM metals are brought together.

**THE METHOD**

To study the optical response of arrays of atomic chains, we employ the TDDFT (Time-Dependent Density-Functional Theory) approach by using the Gaussian 03 code[26] with a B3PW91 hybrid functional[27, 28, 29] and a LanL2DZ basis set[30]. We consider chains comprised of noble metal atoms (Au, Ag and Cu) with varying number of atoms in a chain (2 to 14) and the number of chains in an array (1 to 3). We also investigate chains comprised of Ni atoms, to compare and contrast their properties with those made up of noble metal atoms. Finally, we consider different cases of TM doping (Ni, Pd, Fe, Rh) of pure metallic chains. Unless specified, all inter-atomic distances are set to be 2.89 Å, as in the case of experimentally realized Au chains on NiAl(110) substrate.[14] This choice of bond length is not very critical as we find the results to be not very sensitive to its exact value.

The TDDFT calculations result in the Kohn Sham eigenenergies $E_i$ and the corresponding transition dipole moments $\langle i|D|j \rangle$, which are used to calculate the optical absorption spectrum:

$$A(\omega) = \sum_{i \neq j} (E_i - E_j) |\langle i|D|j \rangle|^2 \, e^{-|E_i - E_j - \omega|/\Gamma}, \tag{1}$$

where ω is the oscillation frequency and Γ is the peak broadening, which we set to 0.2eV. For clarity, in plotting $A(\omega)$ we neglect the contribution of dipole moments with intensity less than 1% of the main plasmon peak. The plasmon peaks are identified as the peaks in the spectrum that emerge when the number of atoms in the chain becomes large enough (typically 6), and for which the magnitude of the spectral peak increases with the number of atoms increasing (indicating the collective/resonance nature of the excitation).

**RESULTS**

In this Section, we consider separately the cases pure Au chains, Au chains doped with TM atoms at different positions, and the case when two pure (noble or TM-atom) chains form a coupled system.

(1) <u>Pure chains:</u> Similar to the case of single chain,[23] increasing the number of atoms in the double chains (2 chains in an array) leads to the generation of a plasmon mode when the number of atoms $n$ reaches a certain threshold value. This mode gains strength and moves to lower energy with further increase in the number of atoms, until $n$ reaches becomes approximately 20 (saturated collective response). Fig.1 shows how this mode is generated when $n \geq 6$ in double Au chains. Importantly, the position of the plasmon peak is not very sensitive to the interatomic distance when it changes between 2Å and 3 Å.

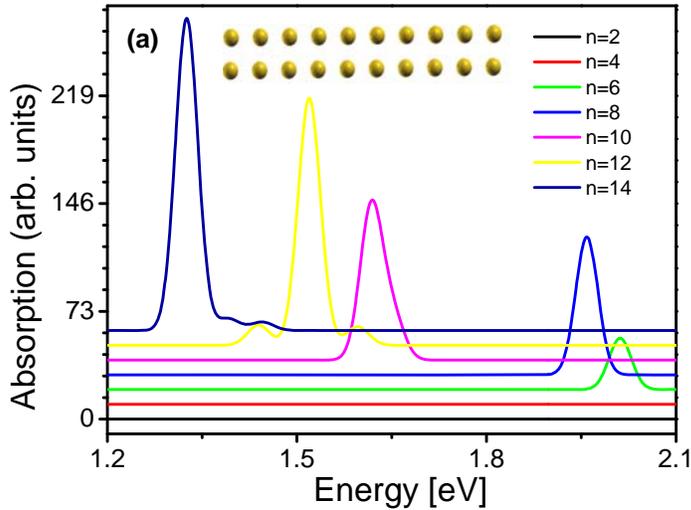

**Figure 1.** Optical absorption spectra of $n$-atom double Au chains at different values of $n$ (an example of the 10-atom double chain is shown on the top of the Figure). Here and in other figures the inter-atomic distance is 2.89Å, following experimental results for Au on NiAl(110).[29]

The absorption spectra of single, double and triple ten-atom Au chains (Fig. 2, left) demonstrate that the plasmon peak moves to the visible range (1.7eV -3.0 eV) as the number of chains in the array increases. A similar effect can be produced by changing the geometry of the system while keeping the number of atoms fixed (Fig. 2, right). We have considered four different geometries, three of which have been previously studied in the case of infinite chains.[31] In the square pattern (SQ3) the two chains are parallel to each other, and each atom coordination number is three. In single zigzag geometry the chains are aligned at some (non - $90^0$) angle to each other, depending on the vertical distance between the two chains. The coordination number for each atom here is four. Examples are ZZ4-eq (the angle is $60^0$, so that the atoms form equilateral triangles) and ZZ4-iso (when the atoms form isosceles triangles). In double zigzag geometry (ZZ3+5) two zigzag chains are combined. Although the average coordination number is four, four atoms in the unit cell alternate between coordination number three and five. The plasmon peaks for SQ3 and ZZ4-iso lies below the visible range. However, in the case of ZZ4-eq and ZZ3+5 chains the peaks are in visible range frequency. Importantly, we used different scales for different graphs in Fig.2 in order to provide a comparison of different systems and focus on the main features of the spectrum in each particular case. So while comparing it should be kept in mind that the strength of the peak in some cases is about three times stronger than other.

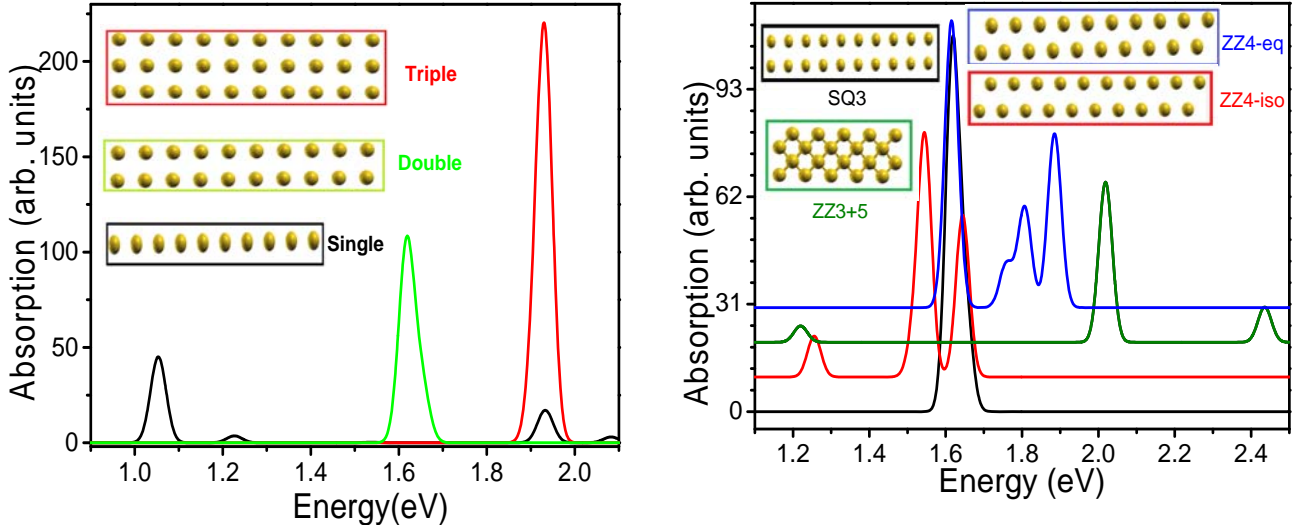

**Figure 2.** Absorption spectra of arrays of pure ten-atom Au chains (left) and pure Au chains with different geometries (right) -- square, zigzag and double zigzag (for details, see the text).

(2) Doped chains: The contrast between the absorption spectra of single and double TM-doped chains is more complicated than that between the corresponding spectra of pure single and double chains. In both cases the spectrum shifts to higher energy and the strength of the plasmon peak increases with increasing number of chains in the array. However, when each chain is doped in the middle with TM atom(s), bringing two chains together increases the number of plasmon modes and moves some of the absorption spectrum peaks into the visible range (Figs.3,4). Although, we present only a few cases with different number of atoms in the chains, similar qualitative picture remains the same in other cases as well. Putting Ni atoms at the end of the chains almost does not affect the plasmon mode of pure Au double chains, whereas Ni in the middle leads to a splitting of the mode into two (Fig. 3). The situation is different in the case of Pd-doped systems (Fig.4): the plasmon peak splits into two around the main peak even when the dopant is at the end of the chain and an additional third mode is observed. With Pd as the dopant, the plasmon peak always splits into two except for the case when we have one Pd atom at the end of one chain.

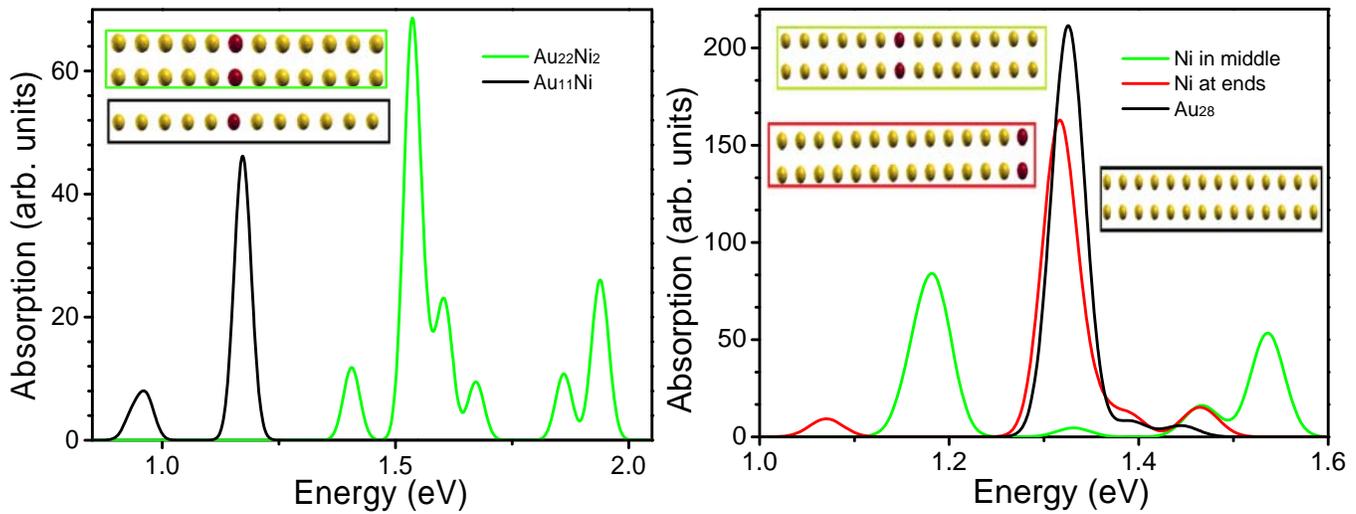

**Figure 3.** Optical absorption spectra of single and double $Au_{11}Ni$ chains doped in the middle (left) and $Au_{28}$, $Au_{26}Ni_2$ (Ni in the middle) and $Au_{26}Ni_2$ (Ni at the end) chains (right).

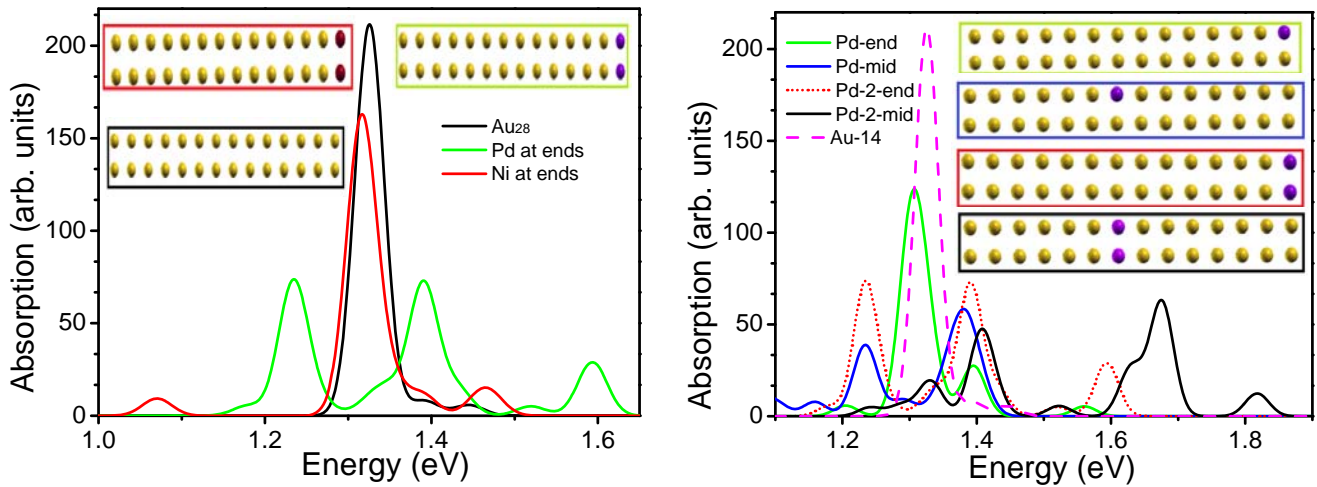

**Figure 4.** Optical absorption spectra of $Au_{28}$, $Au_{26}Ni_2$ (Ni at the end) and $Au_{26}Pd_2$ (Pd at the end) (left Figure) and $Au_{28}$, $Au_{26}Pd_2$ (Pd at ends and in the middle) and $Au_{27}Pd$ (Pd at the end and in the middle) (right Figure) chains

(3) <u>Coupled $M_{14}N_{14}$ chains (M=Au,Ag,Cu; N=Ni, Fe, Pt, Pd, Au, Ag, and Cu)</u>  In the case of double chains, one of which is made up of Au atoms and other either of TM or other noble-metal atoms, the results are also rather nontrivial. In particular, when the Ag chain is coupled with a Cu chain, the resulting plasmon peak is situated halfway between the peaks of pure Ag and pure Cu double chains. On the other hand, when either the Ag or the Cu chain is coupled with an Au chain, the position of the plasmon peak is much closer to that of the pure Au double chain than that of either of the pure Ag or Cu chains, implying that the collective oscillations in the Au chains dominate over the oscillations in two other chains. When the Au chain is coupled to a TM (Ni, Fe, Rh, Pd) chain the plasmon modes are highly suppressed (Fig. 5).

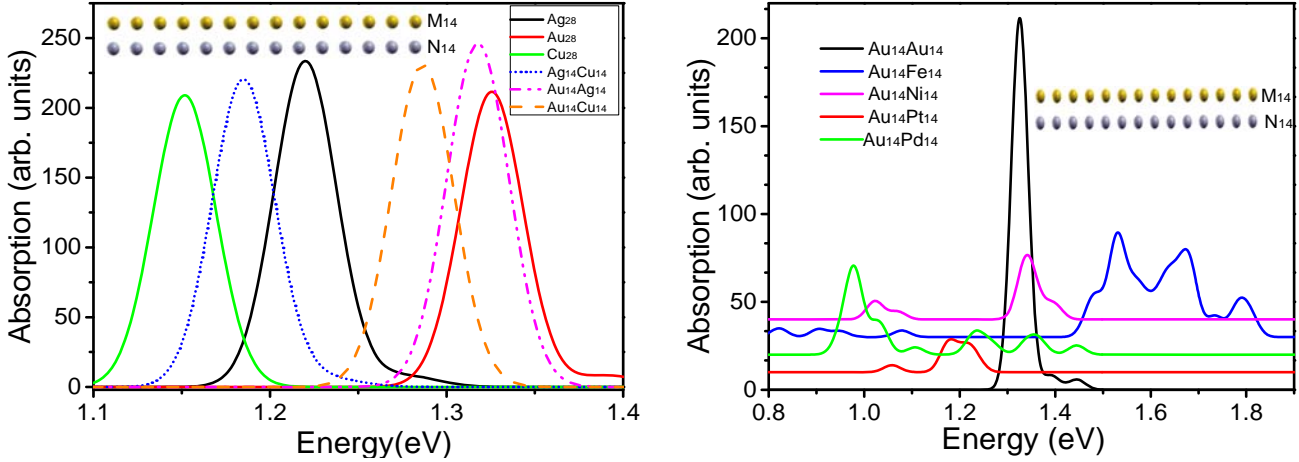

**Figure 5.** The absorption spectra of coupled $M_{14}N_{14}$. Left: $Au_{28}$, $Ag_{28}$, $Cu_{28}$, $Au_{14}Ag_{14}$, $Au_{14}Cu_{14}$, $Ag_{14}Cu_{14}$ chains. Right : $Au_{28}$, $Au_{14}Ni_{14}$, $Au_{14}Pt_{14}$, $Au_{14}Pd_{14}$, and $Au_{14}Fe_{14}$ chains.

**DISCUSSION**

Below we analyze separately the results for three different cases considered in the previous Section.

(1) <u>Pure Chains:</u> As follows from Fig. 2, the position of the plasmon peak in pure chains moves to the higher energies as the number of chains increases. This effect can be described in a simple framework of the jelly model, where the plasmon energy is $\hbar\sqrt{\frac{4\pi n_{el} e^2}{m}}$. In the last expression, $n_{el}$ is the charge density (corresponding to single active s-electron for each atom) and $e$ and $m$ are the electron charge and mass, for which we use the free-electron values. In the multi-chain case, the electron density can be estimated as total number of electrons $nn_{ch}$ (n is the number of atoms in the chain, $n_{ch}$ is the number of chains) divided by the effective volume of the system V. The value for V can be estimated is the volume of the system can be estimated as follows. It is approximately equal to the effective chain length $l$ multiplied by its effective width d and height h. The chain length can be approximated by $l = \Delta l(n-1) + 2R$, where $\Delta l = 2.89$Å is the inter-atomic distance in the chain, and R is the s-orbital radius. The last quantity can be estimated as the radius of the sphere around the atom which contains 90% of the s-electron charge. For Au atoms this gives R=21.07Å. The system height is h=2R in this case, and the system width $d = \Delta l(n_{ch} - 1) + 2R$, which gives:

$$E = \hbar \sqrt{\frac{2\pi n n_{ch} e^2}{mR[\Delta l(n-1)+2R][\Delta l(n_{ch}-1)+2R]}}. \tag{1}$$

Since in a system consisting of only few chains $\Delta l(n_{ch} - 1)$ is much smaller than 2R, and in a system consisting of long chains (n>20), $\Delta l(n - 1)$ is much larger than this quantity, one can use the following approximate expression for the dependence of the plasmon energy on the number of atoms in the chain: $E = \hbar \sqrt{\frac{\pi e^2 n_{ch}}{mR^2 \Delta l}}$. For the parameters mentioned above this gives $E \approx 0.52 eV \sqrt{n_{ch}}$, or 0.52eV, 0.74eV and 0.9eV, for the single-, double- and triple-Au chains, respectively. This is in a reasonable agreement with the results presented in Figs. 1 and 2. One can also estimate the plasmon energy for 2D (infinite) arrays. In this case, $n, n_{ch} \to \infty$, so that one gets from Eq.(1): $E \approx \hbar \sqrt{\frac{2\pi e^2}{mR\Delta l^2}} \approx 1.98 eV$. As it follows from these estimations, indeed the energy of the plasmon peak moves to the visible range as the number of atoms increases.

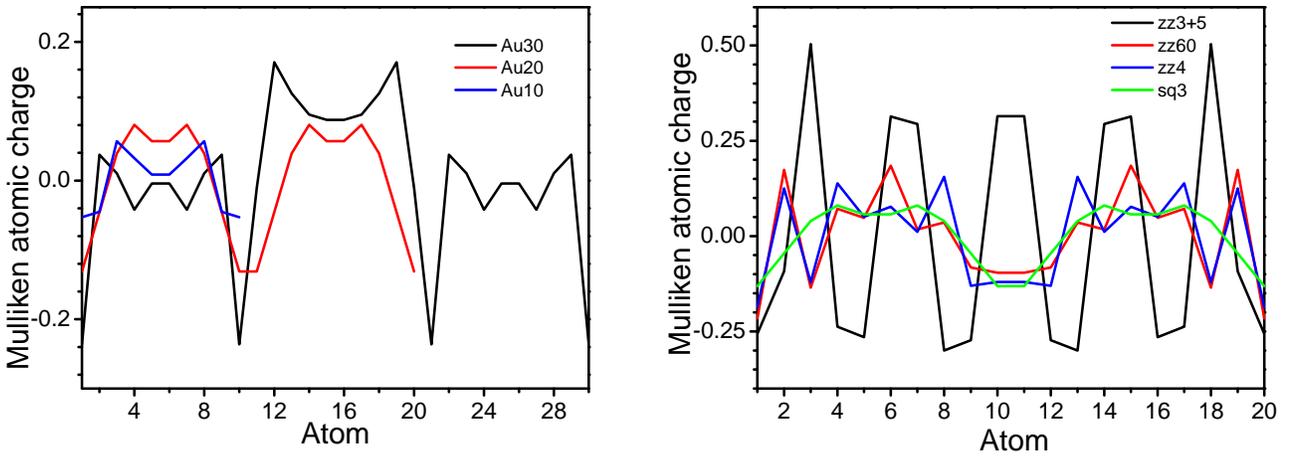

**Figure 6.** Mulliken atomic charge distributions in the pure Au chains (from Fig. 2) in the cases of different sizes (number of atoms) (left) and different geometries (right).

One can also use Eq.(1) to derive an approximate value for the atoms, when the system develops collective plasmon response. Indeed, when $\Delta l(n - 1)$ becomes larger than 2R the plasmon energy weakly depends on the chain length (n's in the numerator and denominator cancel), as should be in the case of the collective behavior of "the extended" system. In other words, an approximate condition of the collective response is $\Delta l(n - 1) > 2R$, or $n > \frac{2R}{\Delta l} + 1$, which gives n>15.5 in our case of Au chains. Taking into account the simplicity of the model, this results is in a rather reasonable agreement with the numerical result n>10. The energy becomes almost n-independent when $\Delta l(n - 1) \gg 2R$, or $n \gg \frac{2R}{\Delta l} + 1 = 15.5$, which is again in agreement with the generally accepted estimation n>20.

(2) <u>Doped Chains:</u> In the case of doped chains, the shape of the spectra strongly depends on the chemical composition of the structure and on the position of the impurity atoms. In the single-chain case, as we have reported earlier,[23] a single impurity leads to an extra peak in the case of some TMs, and the plasmon peaks are suppressed when the number of impurities is large enough (~4-5). The plasmon effects were found to be especially pronounced in the case of Ni, and almost absent in that of Rh. In the case of doped multi-chain structures the situation is more complicated. One would expect in general to find local and collective plasmons in each chain, and that interaction among these many modes leads to extra ("hybridized") peaks, so that the plasmon spectrum is much more complicated and difficult to understand. Moreover, when the chains are close to each other one might anticipate inter-chain oscillations, i.e. the appearance of additional new modes. In order to get some insight on the nature of possible plasmon excitations in different pure multiple Au chains, we present the Mulliken charge distribution in these systems, Fig. 6. As it follows from the Figure, the systems with complex geometry have a non-homogeneous charge distribution, which may be responsible for new local plasmon oscillations. Similarly, strong charge nonhomogeneity was in the case of doped multiple chains, Figs. 7,8 (we do not present results for other cases here).

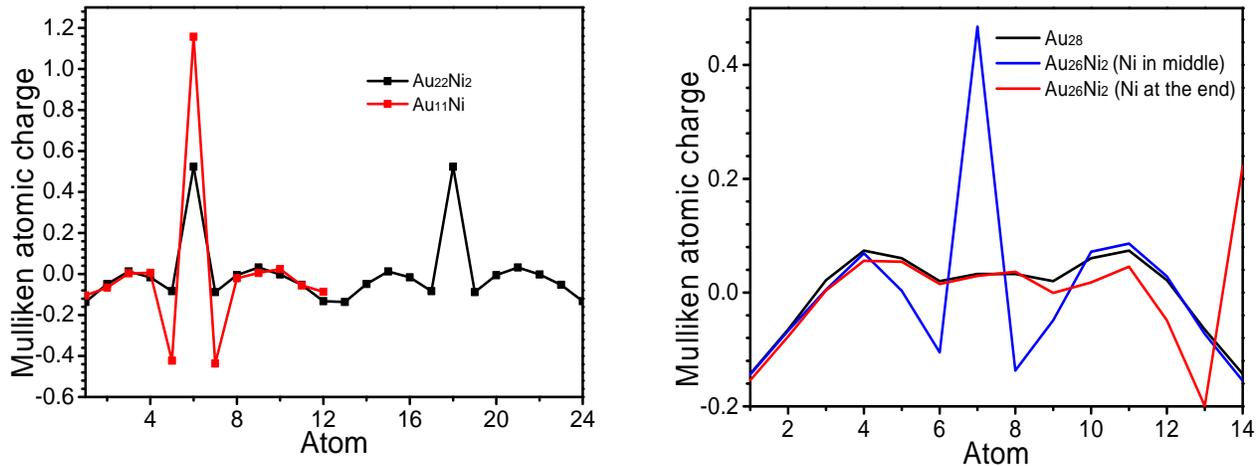

**Figure 7.** Mulliken atomic charge distribution for single and double Au chains with 12 atoms doped with Ni in the middle (left) and the two 14-atom Au chains, pure and doped with one Ni atom in the middle and at the edge (right). On the left figure atoms 13-24 correspond to the atoms 1-12 in the second chain.

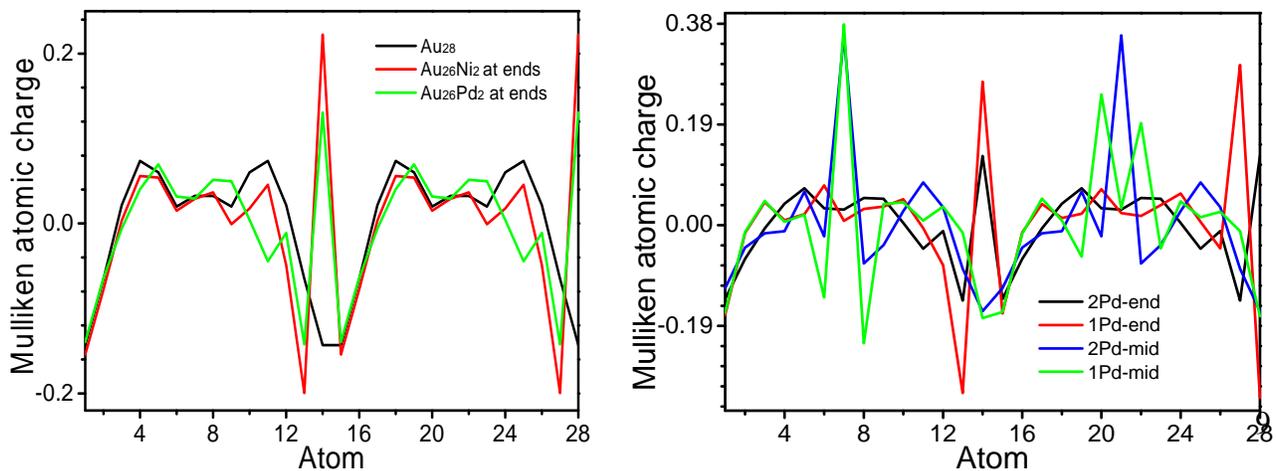

**Figure 8.** Mulliken atomic charge distribution for two 14-atom Au chains: pure and doped with Ni and Pd at the end (left) and doped with one or two Pd atoms in middle or at end (right). Atoms 15-18 correspond to the atoms 1-14 in the second chain.

In the case of multiple chains, generation of hybridized modes from different chains can be explained qualitatively by the following model. The plasmon excitations can be described by a system of coupled equations:

$$i\dot{p}_{i\alpha} = \varepsilon_{i\alpha} p_{i\alpha} + \sum_{ij,\alpha \neq \beta} V_{i\alpha;j\beta}\, p_{i\alpha} p_{j\beta}, \qquad (2)$$

where, $\varepsilon_{i\alpha}$ is the energy of plasmon in chain i, and $\alpha$ is the index of the type of plasmon (local, collective); $V_{i\alpha;j\beta}$ is the inter- and intra-chain plasmon-plasmon interaction. Since the plasmon-plasmon interaction is of the dipole type, it should decay rapidly ($1/R^3$) with increase in the distance R between the plasmon "centers of mass," so only the nearest-chain interaction matrix elements V need to be taken into account (Fig.9).

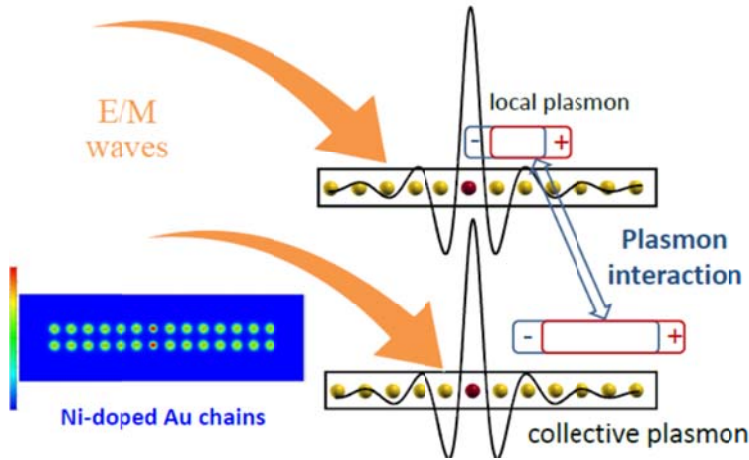

Figure 9. Schematic representation of different types of plasmon excitations and their interactions in double Au chains dped with a TM atom, which can be described by Eq. (2). In this Figure, a local plasmon mode (centered around the right from the impurity potential valley, top chain) interacts with a collective plasmon exitation (on the right half of the bottom chain).

Therefore, the corresponding modes will split, leading to extra modes with respect to the single chain case, as displayed in Figs. 3 and 4, for example. The presence of extra modes in the edge Pd-doped Au chains (Fig.4, the only case for such an edge-doping) can be explained by increased "resonance" oscillations at the end of the chains which interact with collective modes. Though such qualitative analysis can provisionally explain the main feature of the plasmon spectrum in this case, a detailed quantitative analysis is necessary for every particular case.

(3) Mixed Chains: Another surprising result is suppression of the plasmon modes in noble metal chains coupled to a TM chain (Fig. 5). Although both chains support plasmons (though in the TM case the transition magnitude is much weaker), in the coupled case these excitations are damped by a strong mutual scattering between some of the d-wave orbital states of the TM chains oriented "perpendicularly" with respect to the second Au chain with s-orbitals that contribute to the plasmon oscillations. Indeed, as it follows from Fig. 10 for the Mulliken charge distribution for different mixed chains, the in the case of TM chains the system demonstrates in average larger charge localization which may contribute to the plasmon scaterring effects.

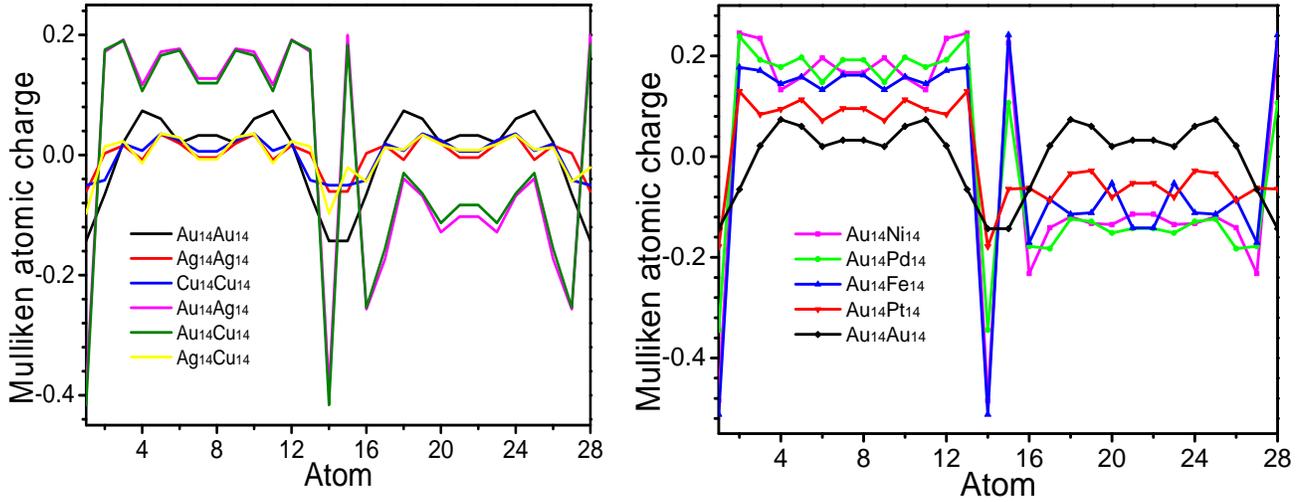

**Figure 10.** Mulliken atomic charge distribution for coupled the $M_{14}N_{14}$ chains (illustrated in Fig. 5) : $Au_{28}$, $Ag_{28}$, $Cu_{28}$, $Au_{14}Ag_{14}$, $Au_{14}Cu_{14}$, $Ag_{14}Cu_{14}$ (left) and $Au_{28}$, $Au_{14}Ni_{14}$, $Au_{14}Pt_{14}$, $Au_{14}Pd_{14}$, $Au_{14}Fe_{14}$ (right). Atoms 15-28 correspond to the atoms 1-14 in the second chain.

**CONCLUSION**

We have analyzed the optical properties of arrays of multiple pure and TM-doped noble metal nanochain systems. In the case of pure chains, the systems demonstrate plasmon peaks when the chain contains about 10 atoms or more. The position of the peak in the optical absorption spectrum moves to lower energies as the number of atoms n increases, and becomes almost n-independent at n>20. The position of the peak moves to higher (visible light) energies when the length of the chain is fixed but number of chains increases. Doping of a single chain with some TM atoms leads to extra plasmon peaks which can be explained by local plasmon oscillations around the potential created by the impurity atom. In the multi-chain case, such a doping leads to multiple in-chain plasmon oscillations, as well as to hybridized modes that consist of modes from the neighboring chains. These extra plasmon modes can be explained as a result of plasmon-plasmon interaction, which leads to the mode splitting. Interestingly, we have found plasmon modes even in pure TM chains. To our knowledge, this is the first report of such an effect in TM materials. Though when such pure chains are coupled to pure noble metal chains, no plasmon modes were found, owing to mutual scattering between s- and d-states excitations in different chains. We have mainly focused on Au chains doped with Ni, Rh or Pd atoms, but as pure Ag and Cu systems behave similarly to pure Au system, one can expect similar picture when Ag and Cu are doped with TM atoms. Provided that such multi-chain systems are built experimentally on a "neutral" substrate, the results obtained in this paper may have applications in different optical technologies, including solar cells and sensors.

ACKNOWLEDGMENTS

The work was supported in part by DOE Grant DE-FG02-07ER46354.